\documentclass[a4paper]{article}

\usepackage{latexsym}
\usepackage{amsmath}
\usepackage{amssymb}
\usepackage{tabularx}
\usepackage{graphicx}
\usepackage{stfloats}
\usepackage{booktabs}
\usepackage{comment}
\usepackage[autolanguage]{numprint}

\usepackage{url}
\usepackage[a4paper,margin=2.75cm]{geometry}

\newcommand{\myparagraph}[1]{\noindent\textbf{#1}\hspace{1ex}}

\DeclareMathOperator*{\argmax}{argmax}
\newcommand{\specialcell}[2][c]{\begin{tabular}[#1]{@{}c@{}}#2\end{tabular}} 

\usepackage{graphicx}

\usepackage{amsmath} 
\usepackage{amssymb} 

\usepackage{stmaryrd}
\usepackage{dsfont}  
\usepackage{accents} 

\usepackage{authblk}
\title{Multimodal Entity Linking for Tweets\footnote{Preprint of Adjali O., Besançon R., Ferret O., Le Borgne H., Grau B. (2020) Multimodal Entity Linking for Tweets. In: Jose J. et al. (eds) Advances in Information Retrieval. ECIR 2020. Lecture Notes in Computer Science, vol 12035. Springer, Cham. The final authenticated version is avalable online at  https://doi.org/10.1007/978-3-030-45439-5\_31 }}
\author[1]{Adjali, Omar}
\author[1]{Besançon, Romaric}
\author[1]{Ferret, Olivier}
\author[1]{Le Borgne, Hervé}
\author[2]{Grau, Brigitte}
\affil[1]{Université Paris-Saclay, CEA, List, F-91120, Palaiseau, France}
\affil[2]{Universit\'e Paris-Saclay, CNRS, LIMSI, 91400, Orsay, France}
\date{}
\begin{document}

\maketitle

\begin{abstract}

In many information extraction applications, entity linking (EL) has emerged as a crucial task that allows leveraging information about named entities from a knowledge base. In this paper, we address the task of multimodal entity linking (MEL), an emerging research field in which textual and visual information is used to map an ambiguous mention to an entity in a knowledge base (KB). 
First, we propose a method for building a fully annotated Twitter dataset for MEL, where entities are defined in a Twitter KB. Then, we propose a model for jointly learning a representation of both mentions and entities from their textual and visual contexts. 
We demonstrate the effectiveness of the proposed model by evaluating it on the proposed dataset and highlight the importance of leveraging visual information when it is available.
\end{abstract}

\section{Introduction}
\label{introductionSection}
Entity linking (EL) is a crucial task for natural language processing applications that require disambiguating named mentions within textual documents. It consists in mapping ambiguous named mentions to entities defined in a knowledge base (KB). To address the EL problem, most of the state-of-the-art approaches use some form of textual representations associated with the target mention and its corresponding entity. \cite{bunescu2006using} first proposed to link a named mention to an entity in a knowledge base, followed by several other local approaches \cite{cucerzan2007large,milne2008learning,dredze2010entity,eshel-EtAl:2017:CoNLL,daher17cicling}  where mentions are individually disambiguated using lexical mention-entity measures and context features extracted from their surrounding text. In contrast, global approaches \cite{ratinov2011local,guo2014entity,pershina2015personalized,globerson2016collective} use global features at the document-level to disambiguate all the mentions within the document and can also exploit semantic relationships between entities in the KB. These approaches have shown their effectiveness on standard EL datasets such as TAC KBP \cite{ji2010overview}, CoNLL-YAGO \cite{hoffart2011robust}, and ACE \cite{bentivogli2010extending} which contain structured documents that provide a rich context for disambiguation. However, EL becomes more challenging when a limited, informal, textual context is available, such as in social media posts. On the other hand, social media posts often include images to illustrate the text that could be exploited to improve the disambiguation by adding a visual context. 

\begin{figure*}[t]
\centering
  \includegraphics[width=0.9\textwidth]{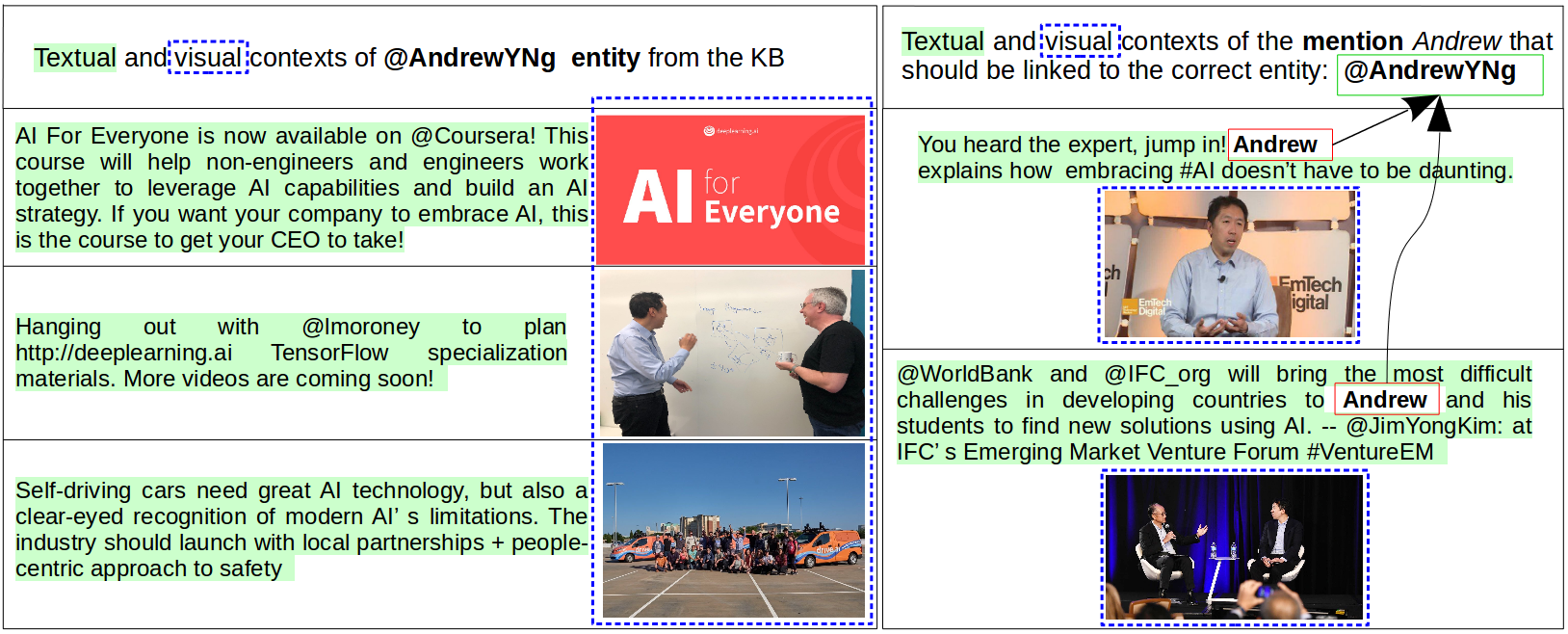}
  \caption{An illustrative example of multimodal Entity Linking on our Twitter dataset. The example  depicts a subset of (text, image) pairs representing the entity @AndrewNg (left), and some example of (text, image) pairs related to the mention \textit{Andrew} (right).}
  \label{exampleMultimodalEL}
\end{figure*}

In this paper, we address the problem of multimodal entity linking (MEL), by leveraging both semantic textual and visual information extracted from mention and entity contexts. We propose to apply  MEL on Twitter posts as they form a prototypical framework where the textual context is generally poor and visual information is available. 
Indeed, an important mechanism in Twitter communication is the usage of a user's screen name (@UserScreenName) in a tweet text. This helps to explicitly mention that the corresponding user is somehow related to the posted tweet. One observation we made is that some Twitter users tend to mention other users without resorting to screen names but rather use their first name, last name, or acronyms (when the Twitter user is an organization). Consider the following example tweet (see Figure~\ref{exampleMultimodalEL}): \begin{quote}\textit{\textbf{Andrew} explains how  embracing AI doesn't have to be daunting}\end{quote} In this example, the mention \textit{Andrew} refers to the user's screen name @AndrewYNg. Obviously, the mention \textit{Andrew} could have referred to any Twitter user whose name includes Andrew. Such a practice, when it occurs, may lead to ambiguities that standard EL systems may not be able to resolve, as most of Twitter users do not have entries in standard knowledge bases. We thus propose to link ambiguous mentions to a specific Twitter KB, composed of Twitter users.

Our contributions are: \textbf{(i)} we investigate the multimodal entity linking task. To the best of our knowledge, this is the first EL work that combines textual and visual contexts on a Twitter dataset; \textbf{(ii)} we construct a new large-scale dataset for multimodal EL 
and above all, define a method for building such datasets at convenience; \textbf{(iii)} we present a new EL model-based for learning a multimodal joint representation of tweets and show, on the proposed dataset, the interest of adding visual features for the EL task. Code and data are available at https://github.com/OA256864/MEL$\_$Tweets\footnote{forked at https://github.com/cea-list-lasti/MEL$\_$Tweets}.

\section{Related Work}\label{sec:related}
\myparagraph{Entity Linking on Twitter}
Recently, several research efforts proposed to meet the challenges posed by the EL task on Twitter media posts. Collective approaches are preferred and leverage global information about tweets in relation with the target mention. \cite{liu2013entity} collectively resolves a set of mentions by aggregating all their related tweets to compute mention-mention and mention-entity similarities while \cite{huang2014collective} takes advantage of both semantic relatedness and mention co-referencing. In a more original way, \cite{shen2013linking} determines the user's topics of interest from  all its posted tweets in order to collectively link all its named entity mentions. Similarly \cite{hua2015microblog} considers social (user's interest + popularity) and temporal contexts. Other collective approaches include in their EL model additional non-textual features. For example, \cite{fang2014entity} and \cite{chong2017collective} use global information of tweets that are  close in space and time to the tweet of the target mention. Finally, \cite{dredze2016twitter} proposes a joint cross-document co-reference resolution and disambiguation approach including temporal information associated with their corpus to improve EL performance. While these works yield interesting results using non-textual features, they often depend on the availability of social data and do not exploit visual information.
 
\myparagraph{Multimodal Representation Learning}
 Joint multimodal representations are used in several multimodal applications such as visual question answering \cite{jabri2016revisiting,agrawal2017vqa}, text-image retrieval \cite{wang2016learning,chami17icmr,chowdhury2018webly} and image captioning \cite{karpathy2015deep,johnson2016densecap} and exploits different approaches e.g., Canonical Correlation Analysis~\cite{tran16cvpr}, linear ranking-based models and non-linear deep learning models to learn projecting image features and text features into a joint space \cite{fukui2016multimodal}. In our work, we take inspiration from \cite{wang2019learning} for our joint representation learning model as it showed satisfying results for multimodal learning. 

To our knowledge, \cite{moon2018multimodal} is the only other work that leverages multimodal information for an entity disambiguation task in a social media context. However, they use a dataset of 12K annotated image-caption pairs from Snapchat (which is not made available), whereas our work  relies on a much larger base (85K samples for the benchmark and 2M to build the KB) and is more in line with the state-of-the-art which mostly uses Twitter as case study for EL on social media. Furthermore, they use Freebase as KB, which does not contain image information and does not allow a multimodal representation at the KB level, leading to a very different disambiguation model.

\section{Problem Formulation}
\label{problemSection}
In Twitter terms, each user account has a unique screen name (@AndrewYNg), a user name (Andrew Ng), a user description (Stanford CS adjunct faculty) and a timeline containing all the tweets (text+image) posted by the user (see Figure~\ref{exampleMultimodalEL}). On this basis, we consider the following definitions:
\begin{itemize}
    \item an entity \textit{e} corresponds to a Twitter user account \textit{u} (generally associated with a person or an organization);
    \item a mention \textit{m} corresponds to an ambiguous textual occurrence of an entity $e_j$ mentioned in a tweet \textit{t} that does not belong to the timeline of $e_j$;
    \item the knowledge base is the set of entities (i.e. Twitter accounts).
\end{itemize}

Formally, we denote the knowledge base $KB=\{e_j\}$ as a set of entities, each entity being defined as a tuple $e_j=(s_j,u_j,TL_j)$ including their screen name $s_j$, user name $u_j$ and timeline $TL_j$ (we did not use the user description in the representation of the entity). The timeline contains both texts and images. A mention $m_j$ is defined as a pair $(w_i,t_i)$ composed of the word (or set of words) $w_i$ characterizing the mention and the tweet $t_i$ in which it occurs: $t_i$ contains both text and images.

The objective of the task consists in finding the most similar entity $e^*(m_i)$ to the mention $m_i$ in the KB according to a given similarity. 
From a practical point of view, we do not compute the similarities between the mention and all entities in the KB. We first select a subset of the entities that are good candidates to disambiguate $m_i$. It is defined as $Cand(m_i)=\{e_j \in KB | w_i \sim u_j\}$ where $w_i \sim u_j$ indicates that the mention words $w_i$ are close to the entity name $u_j$. In our case, due to the nature of the dataset, we only use a simple inclusion (i.e. the mention words are present in the entity name) but a more complex lexical distance that takes into account more variations could be considered~\cite{moon2018multimodal}.
The disambiguation of $m_i$ is then formalized as 
finding the best multimodal similarity measure between the tweet containing the mention and the timeline of the correct entity, both containing text and images:
\begin{equation}
e^*(m_i) =  \argmax_{e_j \in Cand(m_i)}  sim(t_i,TL_j)
\end{equation}

\section{Twitter-MEL Dataset}\label{sec:dataset}
One important contribution of our work is a novel dataset for multimodal entity linking on short texts accompanied by images derived from Twitter posts. The process to build this dataset is mostly automatic and can, therefore, be applied to generate a new dataset at convenience. More details and further analysis on the dataset can be found in~\cite{adjali2020lrec}.

All the tweets and user's metadata were collected using the Twitter official API\footnote{https://dev.twitter.com}. The dataset creation process comprises two phases: the construction of the knowledge base and the generation of ambiguous mentions. As mentioned in Section~\ref{introductionSection}, our multimodal EL task aims at mapping mentions to entities (i.e. twitter users) from our Twitter knowledge base, which are characterized by a timeline, namely the collection of the most recent tweets posted by a given user. 

As a first step, we established a non-exhaustive initial list of Twitter user's screen names using Twitter lists in order to have users that are likely to produce a sufficient number of tweets and be referred by a sufficient number of other users. A Twitter list is a curated set of Twitter accounts\footnote{https://help.twitter.com/en/using-twitter/twitter-lists} generally grouped by topic. From this initial list of users, we started building the KB by collecting  the tweets of each user's  timeline along with its meta-information ensuring that both re-tweets and tweets without images were discarded. 
Moreover, as explained previously, users tend to create ambiguous mentions in tweets when they employ any expression (for example first or last name) other than Twitter screen names to mention other users in their post (see Section~\ref{introductionSection}). Consequently, we have drawn inspiration from this usage to elaborate a simple process for both candidate entity and ambiguous mention generation.

\subsection{Selection of Possibly Ambiguous Entities}
\label{Ambiguous-entity-generation-section}
To ensure that the KB contains sufficiently ambiguous entities, i.e. entities with possible ambiguous mentions, to make the EL task challenging, we expanded the KB from the initial lists with ambiguous entities. 
More precisely, we first extracted the last name from each Twitter account name of the initial list of users. Then, we used these names as search queries in the Twitter API user search engine to collect data about similar users\footnote{Only the first 1,000 matching results are available with the Twitter API.}. 
Users that have been inactive for a long period, non-English users and non-verified user accounts were filtered. 
Furthermore, to ensure more diversity in the entities and not only use person names, we manually collected data about organization accounts. We relied on Wikipedia acronym disambiguation pages to form  groups of ambiguous (organization) entities that share the same acronym.

\subsection{Generation of Ambiguous Mentions} 
After building the KB, we used the collected entities to search for tweets that mention them. The Twitter Search API\footnote{Twitter API searches within a sampling of tweets published in the past 7 days.} returns a collection of relevant tweets matching the specified query. Thus, for each entity in the KB, \textbf{(i)} we set its screen name (@user) as the query search; \textbf{(ii)} we collect all the retrieved tweets; \textbf{(iii)} we filtered out tweets without images. Given the resulting collections of tweets mentioning the different entities of the KB, we systematically replaced the screen name mentioned in the tweet with its corresponding ambiguous mention: last names for \textit{person} entities and acronyms for \textit{organization} entities. Finally, we kept track of the ground truths of each tweet in the dataset reducing the cost of a manual annotation task and resulting in a dataset composed of annotated pairs of text and image.
Although ambiguous mentions are synthetically generated, they are comparable to some extent with real-world ambiguous mentions in tweets. We applied a named entity recognition (NER) system \cite{devlin2019bert} on the tweets of the dataset  which achieved a 77\% accuracy score on the generated mentions. This suggests that these latter are somehow close to real-world named mentions.
\subsection{Dataset Statistics and Analysis}
Altogether, we collected and processed 14M tweets, 10M timeline tweets for entity characterization and 4M tweets with ambiguous mentions (mention tweets) covering 20k entities in the KB. Filtering these tweets drastically reduced the size of our data set. Regarding mention tweets, a key factor in the reduction is the elimination of noisy tweets where mentions have no syntactic role in the tweet text (e.g. where the mention is included in a list of recipients of the message).
Discarding these irrelevant tweets as well as tweets without image left a dataset of 2M timeline tweets and 85k mention tweets. After 3 months of data collection, we found that only 10\% of tweet posts are accompanied by images. In the end, we randomly split the set of mention tweets into training (40\%), validation (20\%) and test (40\%) while ensuring that 50\% of mention tweets in the test set correspond to entities unseen during the training.

\begin{table}
\caption{Statistics on timeline and entity distributions in MEL dataset.}
\label{DistributionTable}
\centering
\small
\begin{tabular}{c@{\hspace{1em}}c@{\hspace{1em}}c@{\hspace{1em}}c@{\hspace{1em}}c@{\hspace{1em}}c}
& \textbf{Mean} & \textbf{Median} & \textbf{Max} & \textbf{Min} & \textbf{StdDev} \\\hline
\specialcell{nb Tweets / timeline (text+image)} &  127.9 & 52  & \numprint{3117} &  1 & 222.2\\
\specialcell{nb ambiguous entities/mention} &  16.5 & 16  & 67 &  2 & 12\\
\end{tabular}
\end{table}

Table~\ref{DistributionTable} shows the timeline tweet distribution of all entities in our KB. As noted by \cite{hua2015microblog}, this distribution reveals that most Twitter users are information seekers, i.e. they rarely tweet, in contrast to users that are content generators who tweet frequently. Along with user's popularity, this has an influence on the number of mention tweets  we can collect. We necessarily gathered more mention tweets from content generator entities, as they are more likely to be mentioned by others than information seeker entities.

\section{Proposed MEL Approach}
\label{approach_section}

Visual and textual representations of mentions and entities are extracted with pre-trained networks. We then learn a two-branch feed-forward neural network to minimize a triplet-loss defining an implicit joint feature space. This network provides a similarity score between a given mention and an entity that is combined with other external features (such as popularity or other similarity measures) as the input of a multi-layer perceptron (MLP) which performs a binary classification on a (mention, entity) pair.
%
\subsection{Features}
\label{features_section}
\myparagraph{Textual context representation} We used the unsupervised Sent2Vec \cite{pgj2017unsup} model to learn tweet representations, pre-trained on large Twitter corpus. We adopted this model as training on the same type of data (short noisy texts) turns out to be essential for performing well on the MEL task (see Sec. \ref{sec:results}). The Sent2Vec model extends the CBOW model proposed by  \cite{mikolov2013efficient} to learn a vector representation of words. More precisely, it learns a vector representation of a sentence \textit{S} by calculating the average of the embeddings of the words (unigrams) making up the sentence \textit{S} and the n-grams present in \textit{S}:
\begin{equation}
V_\textit{S} =  \frac{1}{|R(S)|}\sum\limits_{w \in R(S)} v_w
\end{equation} 
where R(S) is the set of n-grams (including unigrams) in the sentence \textit{S}, and $v_w$ is the embedding vector of the word \textit{w}.
Therefore, the textual context of a mention $m_i$ within the tweet $t_i$ is represented by the sentence embedding vector of $t_i$. We produce then for each mention two continuous vector representations (D=\numprint{700}), a sentence embedding $U^{(i)}_{m}$ inferred using only tweet unigrams and a sentence embedding $B^{(i)}_{m}$ inferred using tweet unigrams and bigrams. Combining their vectors is generally beneficial \cite{pgj2017unsup}. An entity context being represented by a set of tweets (see Section~\ref{problemSection}), given an entity $e_i$, we average the unigram and bigram embeddings of all $e_i$'s timeline tweets yielding  two average embedding vectors $U^{(i)}_{e}$, $B^{(i)}_{e}$  representing the entity textual context used as features.

\myparagraph{BM25 features}
 
Given that the disambiguation task aims at finding the correct entity for a tweet, it can be viewed as an IR problem, where we try to associate a given tweet with the most relevant timeline in the KB. We, therefore, consider, as a baseline, a \textit{tf-idf}-like model to match the mention with the entity: in our case, both the tweet and the timeline are represented as bag-of-words vectors and we used the standard BM25 weighting scheme to perform the comparison.

\myparagraph{Popularity features} Given an entity $e$ representing a Twitter user $u$, we consider 3 popularity features represented by:  $N_{fo}$ the number of followers, $N_{fr}$  the number of friends and $N_{t}$ the number of tweets posted by $u$.

\myparagraph{Visual context features} The visual features are extracted with the Inception\_v3  model  \cite{szegedy2016rethinking}, pre-learned on the 1.2M images of the ILSVRC challenge~\cite{russakovsky2014imagenet}. We use its last layer (D = \numprint{1000}), which encodes high-level information that may help discriminating between entities. For an entity $e_i$, we retain a unique feature vector that is the average of the feature vectors of all the images within its timeline, similarly to the process for the textual context. The visual feature vector of a mention is extracted from the image of the tweet that contains the mention.

\subsection{Joint Multimodal Representation Learning}
\label{multimodal-section}
\begin{figure}[t]
 \centering
  \includegraphics[width=0.7\textwidth]{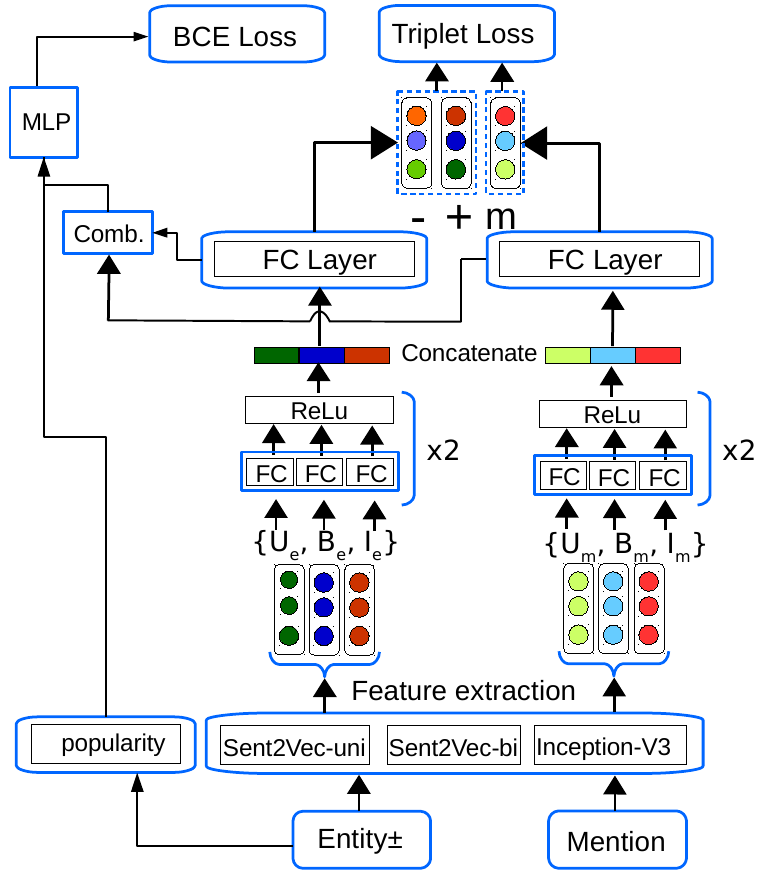}
  \caption{Triplet ranking model for learning entity and mention joint representations.}
  \label{nnArchi}
\end{figure}

The proposed model measures the multimodal context similarity between a mention and its candidate entities. Figure \ref{nnArchi} shows the architecture of the proposed joint representation learning model. It follows a triplet loss structure with two sub-models as in \cite{he2015multi,rao2016noise}, one processing the mention contexts and the other processing the entity contexts. Each sub-model has a structure resembling the similarity model proposed by \cite{wang2019learning}, i.e., it comprises three branches (unigram embedding, bigram embedding, image feature), each in turn including 2 fully connected (FC) layers with a  Rectified  Linear Unit (ReLU) activation followed  by a normalization layer \cite{lei2016layer}. Then, the output vectors of the three branches are merged using concatenation, followed by a final FC layer. Other merging approaches exist such as element-wise product and compact bilinear pooling \cite{fukui2016multimodal,wang2019learning}. However, in our work, simple concatenation showed satisfying results as a first step. Moreover, the mention and entity inputs differ in our task, a mention being characterized by one (text, image) pair and an entity by a large set of (text, image) pairs. Thus, we investigated the performance of our model when the parameters of the two sub-models are partially or fully shared. We found that shared parameters yielded better accuracy results on the validation set, thus the weight of the FC layers of both branches are shared. 

In summary, using the extracted visual and textual features \{$U_m$,$B_m$,$I_m$\}, \{$U_e$,$B_e$,$I_e$\} respectively of a mention $m_i$ and an entity $e_i$, our model is trained to project each modality through the branches into an implicit joint space. The resulting representations are concatenated into $C_m$ and $C_e$  and passed through the final FC layer yielding two multimodal vectors $J_m$ and $J_e$. The objective is to minimize the following triplet loss function:  

\begin{eqnarray*}
\min\limits_{W}\sum\limits_{e^{-} \neq e^{+}} \max(0 , 1 - \lVert f_{m}(m),f_{e}(e^{+})\rVert - \lVert f_{m}(m),f_{m}(e^{-})\rVert ) 
\end{eqnarray*} 
where  \textit{m}, $e^+$ , $e^-$, are the mention, positive and negative entities respectively. $f_m(\cdot)$ is the mention sub-network output function, $f_e(\cdot)$ the entity sub-network function and $\rVert\cdot\lVert $ is the $L_{2}$-norm. The objective aims at minimizing the $L_2$ distance between the multimodal representation of \textit{m} and the representation of a positive entity $p^+$, and maximizing the distance to the negative entity $e^-$ multimodal representation. 
For MEL, we calculate the cosine similarity between the two vectors of a pair ($m_i$, $e_i$) given by $f_m(\cdot)$ and $f_e(\cdot)$ to represent their multimodal context similarity:  $sim(m_i, e_i) = cosine(J^{(i)}_m , J^{(i)}_e)$.

Finally, we propose to integrate the popularity of the entity to the estimation of the similarity by combining the multimodal context similarity score and popularity features through a MLP. Other external features may also be combined at the MLP level, as in our case the BM25 similarity. This MLP is trained to minimize a binary cross-entropy, with label 0 (resp. 1) for negative (resp. positive) entities w.r.t the mention.

\section{Experiments Analysis and Evaluation Setup}\label{sec:eval}
\subsection{Parameter Settings}
We initialized the weights of each FC layer using the Xavier initialization \cite{glorot2010understanding}. Training is performed over 100 epochs, each covering 600k (mention, entity) pairs of which 35k are positive samples. The triplet loss is minimized by stochastic gradient descent, with a momentum of 0.9, an initial learning rate of 0.1 and a batch size of 256. The learning rate is divided by 10 when the loss does not change by more than $1e^{-4}$ during 6 epochs. After 50 epochs, if the accuracy on the validation set does not increase during 5 successive evaluations, the training is early stopped. Regarding the BCE branch, the MLP has two hidden layers with one more neuron than the number of inputs and $tanh$ non-linearities. The BCE loss is minimized using L-BGFS with a learning rate of $10^{-5}$. The network is implemented in PyTorch \cite{paszke2017automatic}.

Since our approach is not directly comparable with previous works \cite{moon2018multimodal}, we compare the results with different configurations and baselines. As 50\% of mentions are unique in each split of our dataset, we do not take the standard most frequent entity prediction as a baseline, which links the entity that was seen most in training to its corresponding mention. We rather consider the popularity features and standard textual similarity measures (BM25), presented in section~\ref{features_section}.
Moreover, for comparison sake, we also report in Table~\ref{results_table} the performance of
a baseline combination of the features using an Extra-Trees classifier~\cite{geurts2006extremely}, compared with our Joint Multimodal Entity Linking approach (JMEL), which combines the textual and visual features at the representation level. 
Table~\ref{features-description_Table} summarizes the features and the combination models used in these experiments.

\begin{table*}[t]
    \caption{Features and models used in our experiments.  \label{features-description_Table}}
  \centering
  \begin{tabular}{@{}p{0.2\textwidth}p{0.8\textwidth}@{}}
   Features & Description (see Section~\ref{features_section}) \\
   \toprule
   Popularity (Pop)	&  Baseline feature where the most popular entity is selected; \\
   BM25 &  Standard textual context similarity with BM25 weighting;   \\
   S2V-uni &  Similarity measured between the unigram embeddings extracted using the Sent2Vec language model; \\
   S2V-bi & Similarity measured between the bigram embeddings extracted using  the Sent2Vec language model;\\
   S2V & For easy readibility, we use the S2V notation to represent the combination of S2V-uni and S2V-bi;\\
   Img & Similarity measured between the image features extracted using  the pretrained Inception-V3 model; \\
   ET(\textit{X}) & Combination of features \textit{X} using an Extra-Trees classifier;\\
   JMEL(\textit{X}) & Combination of features \textit{X} with our joint multimodal representation model (see Section~\ref{multimodal-section}). \\

  \end{tabular}
\end{table*}

In order to more thoroughly assess the contribution of textual and visual information in the MEL task, we compare the performance of the proposed model when the visual features are combined with sentence embeddings that are derived from different models the following notable models: 
Skip-Thought \cite{kiros2015skip} (D=\numprint{4800}) trained on a large corpus of novels, Sent2Vec \cite{pgj2017unsup} unigram and bigram vectors (D=700) trained respectively on English Wikipedia and on a Twitter corpus, InferSent \cite{conneau-EtAl:2017:EMNLP2017} (D=\numprint{2048}) Bi-LSTM+max pooling model trained on the Standford SNLI corpus, BERT \cite{devlin2019bert} trained on BooksCorpus \cite{zhu2015aligning} and on English Wikipedia and ELMo \cite{peters2018deep} trained  on the One Billion Word Benchmark \cite{chelba2014one}. For BERT and ELMo, sentences are represented by the average word embedding.
\begin{table*}[tp]
    \caption{Multimedia Entity Linking results (accuracy). \label{results_table}}
  \centering
  \begin{tabular}{l@{\hskip 2em}c@{\hskip 2em}c}
    & Valid.  & Test \\
    \toprule
    Single features\\
   \midrule
   Popularity	& 0.369	& 0.590 \\
   BM25	& 0.415	& 0.433 \\
   S2V-uni  &  0.482 & 0.513 \\
   S2V-bi  &  0.487 & 0.523 \\
   Img  &  0.290 & 0.299 \\
   \midrule
    Combination of features with an ExtraTrees Classifier\\
   \midrule
   ET(S2V)  &  0.495 & 0.529 \\
   ET(S2V + Img)   & 	0.507 &	0.542 \\
   ET(S2V + Img + Pop) & 0.585 & 0.627 \\
   ET(S2V + Img + Pop + BM25) & 0.654 & 0.671\\
   \midrule
    Combination of features with our JMEL model\\
   \midrule
   JMEL(S2V) & 0.628 & 0.724 	 \\
   JMEL(S2V + Img)  & 0.639	 &	0.731 \\
   JMEL(S2V + Img + Pop) & \textbf{0.767} & \textbf{0.776} \\
   JMEL(S2V + Img + Pop + BM25) & \textbf{0.795} & \textbf{0.803} \\
    \bottomrule
  \end{tabular}
\end{table*} 

\subsection{Results} \label{sec:results}
Table~\ref{results_table} reports the accuracy results on the validation and test sets for the binary classification task about the correctness of the first entity selected from the KB for a given mention. First, we note that the baseline points out an imbalance in our dataset, as 59\% of the mentions in the test set correspond to the most popular entity among the candidate entities, compared to 36.9\% in the validation set. Note that for some Entity linking datasets, popularity can achieve up to 82\%  accuracy score \cite{dai2018entity}. We also observe that our popularity baseline outperforms the combination of textual and visual features with the Extra-Trees classifier: this indicates that the features extracted from the textual and visual contexts, when naively used and combined, produce poor results. In contrast, our model achieves significant improvements on both the validation and test sets compared with the popularity baseline and the Extra-Trees combination. We see that combining additional features in the JMEL model (popularity and BM25) also provides significant performance gain. Regarding the visual modality, although considering it alone leads to poor results, its integration in a global model always improve the performance, compared to text-only features.

\myparagraph{Image and sentence representation impact analysis} Table \ref{tab:res_embedding_eval} reports the MEL accuracy on the validation and test sets with various sentence representation models. It shows that the integration of visual features always improves the performance of our approach, whatever the sentence embedding model we use, even though the level of improvement varies depending on the sentence model. For example, while the results of averaging BERT word embeddings and InferSent are comparable using images on the test set, InferSent performs significantly better than BERT using text only. This emphasizes the role of the visual context representation of mentions and entities to help in the EL task. If we look at the performance of the various sentence embeddings models, we can see that 
the sent2vec model trained on a Twitter corpus outperforms all other embeddings: this reveals the importance of training data in a transfer learning setting. Indeed, we observe that the sent2vec model trained on English Wikipedia produces worse results than the model trained on the target task data (Twitter). Hence, we can assume that other models that achieve good results when trained on a general corpus (such as InferSent or BERT) would  get better results if trained on a Twitter collection.

\begin{table*}
  \caption{Impact of sentence embeddings on EL results.}\label{tab:res_embedding_eval}
  \centering
  \begin{tabular}{lcc@{\hskip 2em}cc}
   & \multicolumn{2}{c}{Valid} & \multicolumn{2}{c}{Test} \\
  Sent. Embedding & Txt & Txt+Img & Txt & Txt+Img \\
  \hline
  S2V-uni(Twitter) &  0.592 & \textbf{0.611} & 0.698 & \textbf{0.708} \\
  S2V-uni(Wiki) & 0.499 &  \textbf{0.538} & 0.625 & \textbf{0.654} \\
  S2V-bi(Twitter) & 0.616 & \textbf{0.637} & 0.709 & \textbf{0.716} \\
  S2V-bi(Wiki) & 0.511	& \textbf{0.547} & 0.639 & \textbf{0.663}  \\
  InferSent(GloVe) & 0.559 & \textbf{0.579} & 0.666 & \textbf{0.683}  \\
  InferSent(fastText \cite{joulin2016fasttext})  & 0.551 & \textbf{0.570} & 0.671 & \textbf{0.689} \\
  Avg. BERT	& 0.580 & \textbf{0.594} & 0.641 & \textbf{0.687} \\
  Avg. ELMo & 0.524 & \textbf{0.563} & 0.605 & \textbf{0.655} \\
  Skip-Thought & 0.464 & \textbf{0.511} & 0.575 & \textbf{0.605}  \\
  S2V(Twitter) & 0.628 & \textbf{0.639} & 0.724 & \textbf{0.731} \\
  S2V(Wiki)   & 0.524 & \textbf{0.551} & 0.652 & \textbf{0.666}\\
  \hline
  \end{tabular}
\end{table*}

\noindent\textbf{Error analysis}\hspace{1em} 
We identified several potential sources of errors that may be addressed in future work. First, in our approach we characterize entity contexts with a collection of text/image pairs and by taking their mean, we consider all these pairs equally important to represent an entity. However, by manually checking some entities, we note that each timeline may contain a subset of outlier pairs and more specifically, images that are not representative of an entity. Sampling strategies may be employed to select the most relevant images and to discard the misleading outliers. Moreover, our model fails on some difficult cases where the visual and textual contexts of entity candidates are indistinguishable. For example, it fails on the mention "\textit{post}", by linking it to the entity \textit{@nationalpost} instead of \textit{@nypost}, two entities representing news organizations whose posts cover various topics. One additional bias is that we restricted our dataset and KB to have only tweets with images. Meanwhile, we observed that tweets without images tend to have more textual context. Thus, it would be interesting to include tweets without images for further experiments.

\section{Conclusion}
We explore a novel approach that makes use of text and image information for the entity linking task applied to tweets. Specifically, we emphasize the benefit of leveraging  visual features to help in the disambiguation. We propose a model that first extracts textual and visual contexts for both mentions and entities and learns a joint representation combining textual and visual features. This representation is used to compute a multimodal context similarity between a mention and an entity. Preliminary experiments on a dedicated dataset demonstrated the effectiveness of the proposed model and revealed the importance of leveraging visual contexts in the EL task. Furthermore, our work is all the more relevant with the emergence of social media, which offer abundant textual and visual information. In that perspective, we propose a new multimodal EL dataset based on Twitter posts and a  process for collecting and constructing a fully annotated multimodal EL dataset, where entities are defined in a Twitter KB. 

Further exploration is still needed concerning certain points in our model: in particular, our future work includes  exploring attention mechanisms for text and images  
and experimenting different sampling strategies for the triplet loss.

\paragraph{Acknowledgments.}
This research was partially supported by Labex DigiCosme (project ANR11LABEX\-0045DIGICOSME) operated by ANR as part of the program ``Investissements d'Avenir'' Idex Paris Saclay (ANR11IDEX000302). 

\bibliographystyle{plain}
\bibliography{mael-ecir.bib}

\end{document}